# First principles calculations of Structure and electrostatic properties of non-ligated CdSe nanoclusters

Anteneh G Tefera[1], Mogus D Mochena[1], Elijah Johnson[2] and James Dickerson[3]
[1]Department of Physics, Florida A & M University, Tallahassee, FL 32307
[2]School of Environment, Florida A & M University, Tallahassee, FL 32307
[4]Department of Physics, Brown University, Providence, RI 02912

Structural and charge relaxation of nanoclusters of CdSe of diameter 1 – 2 nm are studied with first principle calculations. The relaxations cause significant distortions of smaller systems of ~ 1 nm in diameter and have very minimal effect on the larger systems of ~ 2 nm in diameter. The Cd atoms are pulled in while the Se atoms are pulled out, which results in the flattening of Cd-terminated surface and retention of a zig-zag surface for Se-terminated surface. The surfaces terminated with both Cd and Se result in significant geometrical distortion resulting in charge concentrations at the distorted sites. The associated dipole and quadrupole moments are a function of the distortion or unequal distribution of Cd and Se. The HOMO and HOMO-1 orbitals are located on or near the distortions. Based on the tetrahedral coordination and robustness of the core structure to surface relaxation, the ~ 2 nm–diameter NC is the best candidate for building macroscopic structures from NCs.

## I. Introduction

Semiconductor nanocrystals (NCs) have attracted a lot of interest in the last two decades mainly because of the size-dependent tunability of their electronic spectrum, which results in many electronic, optical and optoelectronic applications.[1] Cadium selenide-based NCs are among the most widely studied nanomaterials due to the ease of their synthesis with a high degree of control by hot – injection technique first developed by Murray et al.[2] In this colloidal synthesis technique, the inorganic core of CdSe is typically covered with organic capping ligands that facilitate the growth of the nanoparticles with desired size and shape by controlling the kinetics of the growth. The ligands solubilize the inorganic cores; prevent them from coagulating with each other, and control their interactions with the solvent. While the ligands play advantageous role in some aspects, they also introduce complexity not only to the structure but also to the nature of the interactions and properties. In particular, the ligand exchange mechanism of such hybrid structures is not fully understood.[3] NMR, XPS, and EXAFS experiments have been performed to determine the nature of the interactions at the surface.[4,5] The studies, however, barely scratch the complexity. Computationally, even a moderately sized hybrid nanostructure (~ 2 nm) poses a daunting challenge to first principles calculations, such as time dependent density functional theory, which are often used to compute the excitation spectrum for optical studies.[6]

For device applications, the colloidal suspensions that are comprised of the core–ligand system is often cast into thin films. Such films, fabricated from of these hybrid NCs, are insulators due to the organic ligands despite the semiconducting cores.[1,7] Postmortem techniques are applied to remove the ligands in order to obtain the desired functionality.[8,9] Such protocols are complicated. Recently ligand–free synthesis techniques have been reported. Kadlag et al have synthesized negatively charged CdS NCs that could sustain colloidal dispersion through electrostatic repulsion with a nonstoichiometric technique that leads to excess sulfide ions.[10] Holman and Kortshagen have reported synthesizing nanocrystal inks of germanium without ligands by solubilizing the bare Ge NCs in select solvents.[11] It is assumed that the resulting colloidal dispersion is sustained with electrostatic interaction as well. Further, increased interest in depositing colloidal NCs into mesoporous thin films such as that of $TiO_2$, has arisen where electrostatic interactions play a significant role.[12]

Despite their importance, unlike the electronic and optical properties, the electrostatic properties of NCs have received very little attention. Rabani et al studied the electronic and electrostatic properties of CdSe NCs using empirical pseudopotential method.[13] Rabani also studied passivated CdSe using classical molecular dynamics and pointed out that the electrostatic field inside a nanocrystal can affect significantly the optical and charging energies as well as the self-assembly of the nanoclusters due to the long range interaction.[14] The motivation for this work arises partly from the need to understand the electrophoretic deposition of NCs whereby the colloidal NCs are driven by electric field to electrodes suspended in polar or non-polar solvents to form thin films. The intent of this work is to 1) determine the size range in which CdSe clusters begin forming enough tetrahedral bonds in nearly spherical NCs so as to form a closely packed bulk structure 2) find out the associated electrostatic properties of the NCs. We will not consider interactions between the NCs suspended in a dielectric medium as it is beyond the plane wave based *ab initio* molecular dynamics package or the small frozen core approximation of the all–electron package used in our computations. Nevertheless the results can be taken as a first step for the non-ligated, colloidal NCs that are stabilized through electrostatic interactions[10,11] despite the fact the clusters used in this work are charge neutral overall. The rest of the paper is organized as follows: in section II, we present the computational method, followed by results and discussion in section III and conclusion in section IV.

## II. Computational details

A systematic search for local minima on the potential energy surface of large clusters is computationally too costly.[15] Therefore, an alternative approach is used to determine the ground state structures by cutting a section of the bulk structure and optimizing it. Small clusters of atoms do not have enough atoms to bond with coordination numbers similar to those of bulk structure. Our previous studies on clusters of GaAs indicate that some atoms of the clusters, when they reach the size of approximately 36 - 40 atoms, have bulk like coordination.[16] Therefore, it is reasonable to assume that atoms of NCs of 1 nm diameter consisting approximately of 55 atoms, except those on the surface, have coordination similar to those of their bulk counterparts. Therefore, we cut out nearly spherical clusters of CdSe from wurtzite bulk structure by setting the origin at the center of the supercell since centering on either Cd or Se results in unequal numbers of Cd and Se.[17] Wherever single bonds are present, they were removed from the surface as in ref. 14 and only atoms with at least 2 bonds on the surface are allowed. The input configurations so prepared were optimized in two different ways with Vienna ab-initio simulation package (VASP) to determine the ground state structures.[18-20] In the first case, the whole structure was optimized from the outset with VASP without taking any symmetry into account; the resulting optimized structure will be termed A for brevity in the rest of the paper. In the second case, the symmetry of the structure was taken into account during optimization and the structure will be termed B. The structure with the lowest energy from the two optimizations was taken as the reference lowest energy structure.

We performed the computations with the projector augmented wave (PAW) psedopotentials with energy cut off of 275 eV using the generalized gradient approximation of Pedrew and Wang (PW91) for the exchange correlation function.[21] The supercell computations were generated in reciprocal space at gamma point. A vacuum region of 12 Å was placed around the cluster to avoid interaction with periodic images. A linear mixing of input and output charge densities in the Pulay scheme was used during the self consistency loops. To ensure the lowest energy structures are indeed the structures with the lowest energy, we performed an additional test. We re-optimized the output structures with Molecular Orbital Package (MOPAC),[22] a classical molecular dynamics code as implemented in Amsterdam Density Functional (ADF) Package,[23] and then with VASP again. None of the structures resulted in lower energy. The optimized structures were then analyzed with VESTA code [24] to look at the bond relaxations and with ADF analysis software to study the charge distribution, moments, highest occupied molecular orbitals (HOMO) and HOMO-1. The latter required optimizing the lowest energy VASP structure again with ADF DFT code. The energies of the triple zeta polarized basis set at small frozen core level for $Cd_{33}Se_{33}$ and $Cd_{45}Se_{45}$ agreed with those obtained from VASP within 0.8%. Therefore, all the output VASP geometries were run with ADF using single point calculation with triple zeta polarized basis set at small frozen core level to generate the output for analysis.

## III. Results and Discussion

Würtzite crystals are stacks of hexagonal planes of atoms and have the highest packing ratio of 74%, like their closely related face centered cubic crystals. The question we want to address is what is the right size of spherical NCs that could lead to such a bulk structure? There are two competing issues that must be taken into account while building macroscopic structures from microscopic units. The first is the coordination number of the atoms of the clusters and the other is the interstitial space between the spherical NCs. Obviously, the larger the spherical volume is the larger the interstitial space it creates as the spheres form a close packed structure. On the other hand, to have as small an interstitial space as possible is preferable to get a more dense structure, which means having as small a cluster as possible. The smaller the structure is, however, the more surface atoms it has, that is, fewer number of atoms with bulk-like coordination. These two competing aspects must be taken into account while building a macroscopic structure from microscopic units. In the case of a CdSe cluster, it must have as many atoms with tetrahedral coordination as possible within the interior region. We define the interior region in this work as the region where atoms of Cd or Se have four bonds and the surface as the region where Cd or Se have less than 4 bonds. To find out the optimal size of NCs with the desired properties, we optimized würtzite NCs of varying diameters, $Cd_{33}Se_{33}$ (~1.3 nm), $Cd_{45}Se_{45}$ (~1.5 nm), $Cd_{72}Se_{72}$ (~1.7 nm) and $Cd_{90}Se_{90}$ (~2 nm).

### III-1. $Cd_{33}Se_{33}$

$Cd_{33}Se_{33}$ was determined to be a magic structure using mass spectral studies by Kasyuya et al.[25] In Fig. 1, the input structure is presented; the interior atoms of Cd (pink) and Se (blue) are colored differently from the surface atoms of Cd (red) and Se (green). The interior atoms constitute 36% of the total, which shows that most of the atoms are surface atoms. These surface atoms are triply and doubly coordinated, with the latter residing at the corners of the (0001) and $(000\bar{1})$ surfaces and on $(11\bar{2}0)$ surface. The stacking in z direction (or c-axis) consists of planes of Se and Cd atoms resulting in unequal ratio of Cd and Se atoms at the uppermost and lowermost planes. The two planes between the upper and lower ones have equal numbers of Cd and Se atoms, however, each type of atom is distributed in reverse order from the other in $[11\bar{2}0]$ direction. The $(11\bar{2}0)$ and (1120) surfaces have equal numbers of Cd and Se atoms with no net charge, however, due to the way Cd and Se atoms are distributed will have quadrupole moments. In addition, the $(11\bar{2}0)$ is very jagged, which will affect the geometry and the charge distribution significantly. So how the NCs are terminated, i.e. what radius and what kind of surface distribution of surface atoms, is of crucial importance.

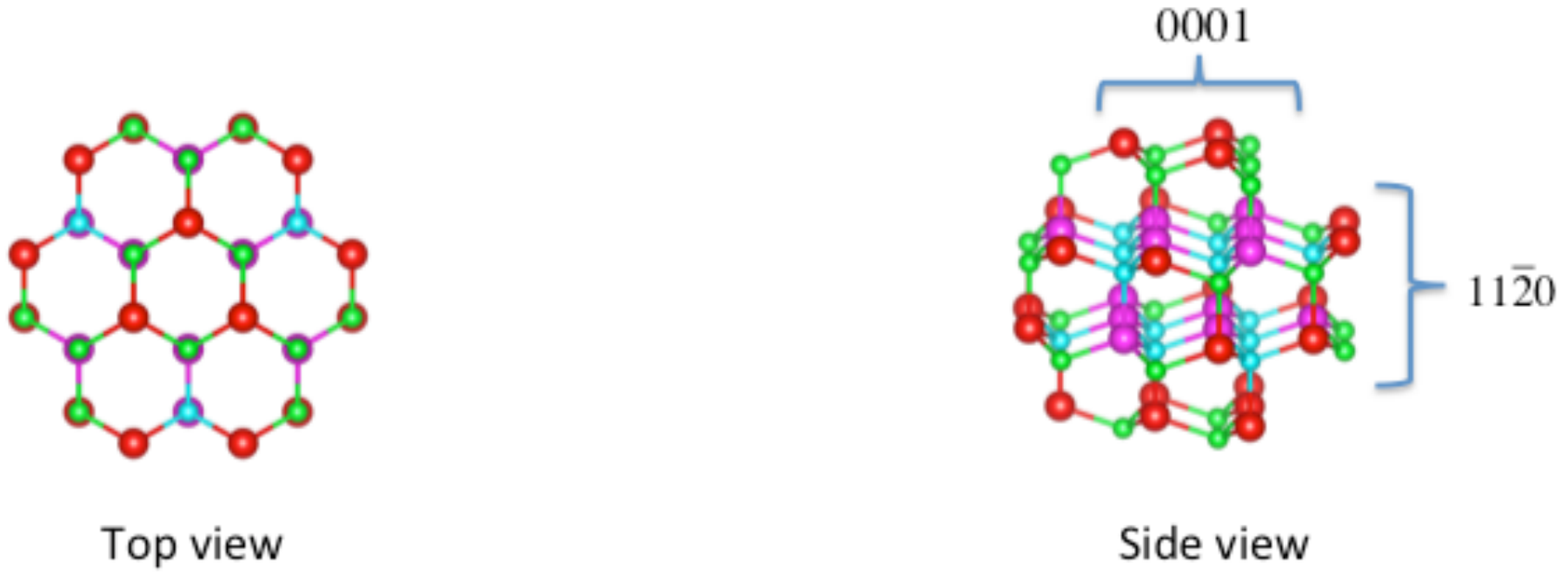

Fig. 1. $Cd_{33}Se_{33}$ input structure. Pink (Cd) and Blue (Se) are interior atoms.

The output structures were first constructed with the ball and stick model using two different softwares. In either case, we noticed that there were a substantial number of Cd-Cd bonds. To make sure these bonds were actually formed, we performed Bader charge analysis to determine the bond paths and the charge density distribution. We present in Fig. 2 one such distribution for structure A. Similar Bader charge analysis was performed for all the optimized structures in this work to check the presence of Cd-Cd bonds; [26] and no Cd-Cd bonds were found. Therefore, we examined only Cd-Se bond formations subject to the constraint of non-formation of the Cd-Cd bonds.

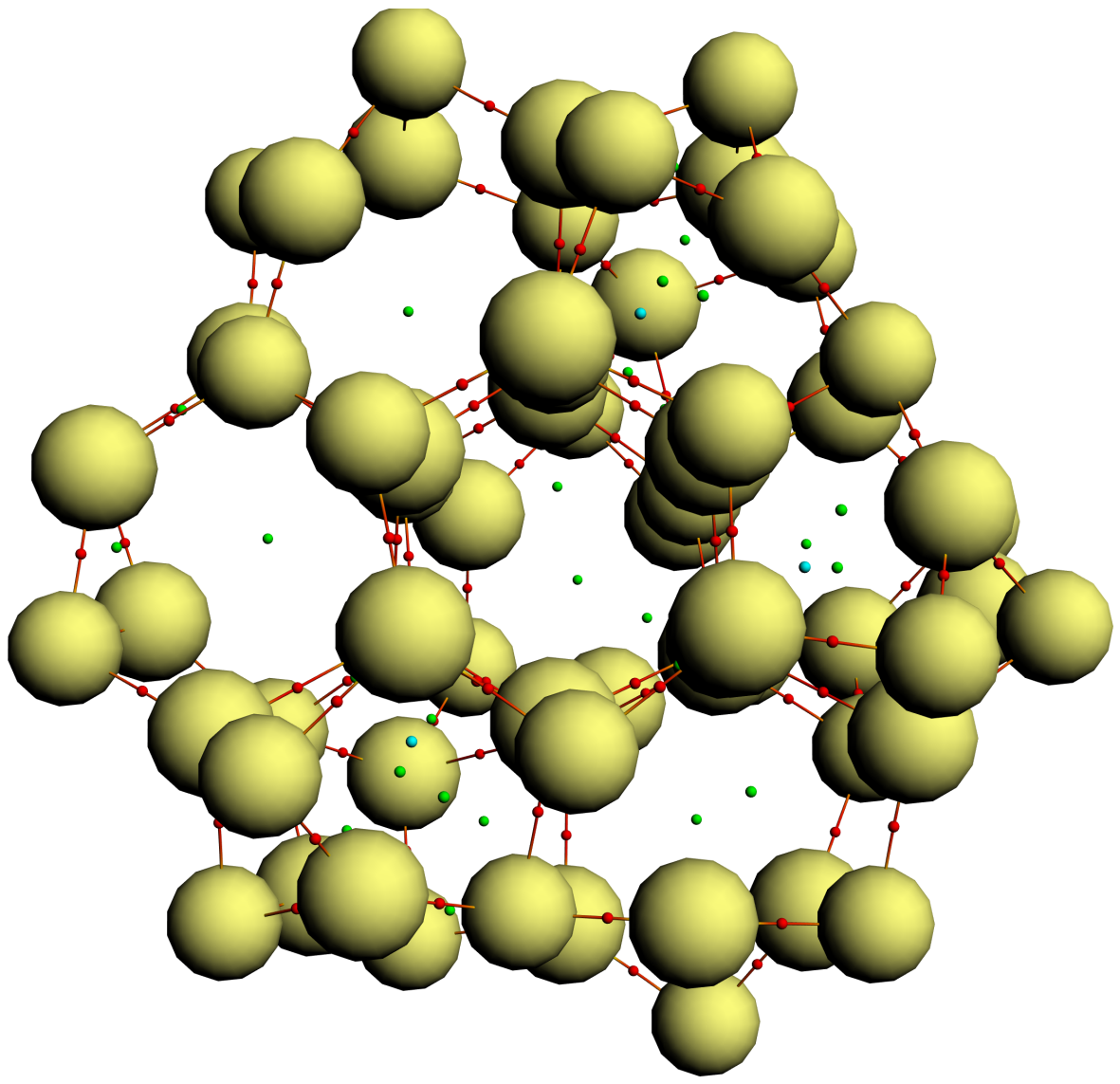

Fig. 2. Top view. The balls represent charge densities, the red lines represent bond paths, the red dots are the bond critical points, and the blue dots are the ring critical points.

One way to ascertain relaxation is to look at its effect on bond distances. The output structures A and B were analyzed by varying the bond lengths with VESTA. For each structure, we started out with the Cd-Se bond length of 2.64 Å for the w*ü*rtzite structure, and varied the length until we obtained the same or approximately same number of bonds as in the input structure. It turns out this bond length is less than the critical bond length at which the Cd-Cd bond appears, which is approximately one Å less than its un-relaxed length of 4.30 Å. In Fig. 3, we present results for the two $Cd_{33}Se_{33}$ structures. To see the effect of symmetry on relaxation, we optimized the structure without and with symmetry constraints imposed. In the bottom rows, we present the results for structure A. At the bond length of a = 2.64 Å, 42 Cd-Se bonds are left, 41% of the 102 Cd-Se input bonds. The relaxation is not confined to surface regions, but also to the interior. Chains of atoms are formed mostly on the surface. As the bond length is increased, more bonds were formed and at a = 2.85 Å, the number of bonds increased to 102, the same number of bonds as that of the input structure. Further increase of the bond length only results in relatively small number of bonds. These relaxations reduced the energy by 0.5 eV per Cd-Se pair, slightly less than the 0.7 eV reported by Puzder et al.[27] It is also seen from Fig. 3 that the Cd-Se bond length has increased while the Cd-Cd bond length (not drawn) has decreased on the average. This result is consistent with that of Puzder et al. The coordination is affected by the strong relaxations, particularly at the corners of Fig. 3 side view. As was reported previously,[28] the doubly coordinated atoms become triply coordinated. However, three doubly coordinated atoms (which became two at a=3.0 Å) remain at $(\bar{1}010)$ surface. The effects of the relaxations are 1) the Cd atoms on (0001) surface are pulled in such that the surface becomes flat consisting of Cd and Se atoms, instead of the zigzag surface resulting from the tetrahedral bonding 2) the atoms on the jagged corners (referred to side view in Fig. 3) form large hexagonal rings as sharp edges are being smoothed out during optimization 3) The inner atoms mostly maintain their tetrahedral coordination except for the ones closer to the surface, which as they are pushed in (Cd) and out (Se) reduce to triply coordinated atoms. Overall, hexagonal rings of the input geometry are distorted, in particular, the ones on or near the $(11\bar{2}0)$ and (1120) as should be expected from the input geometry.

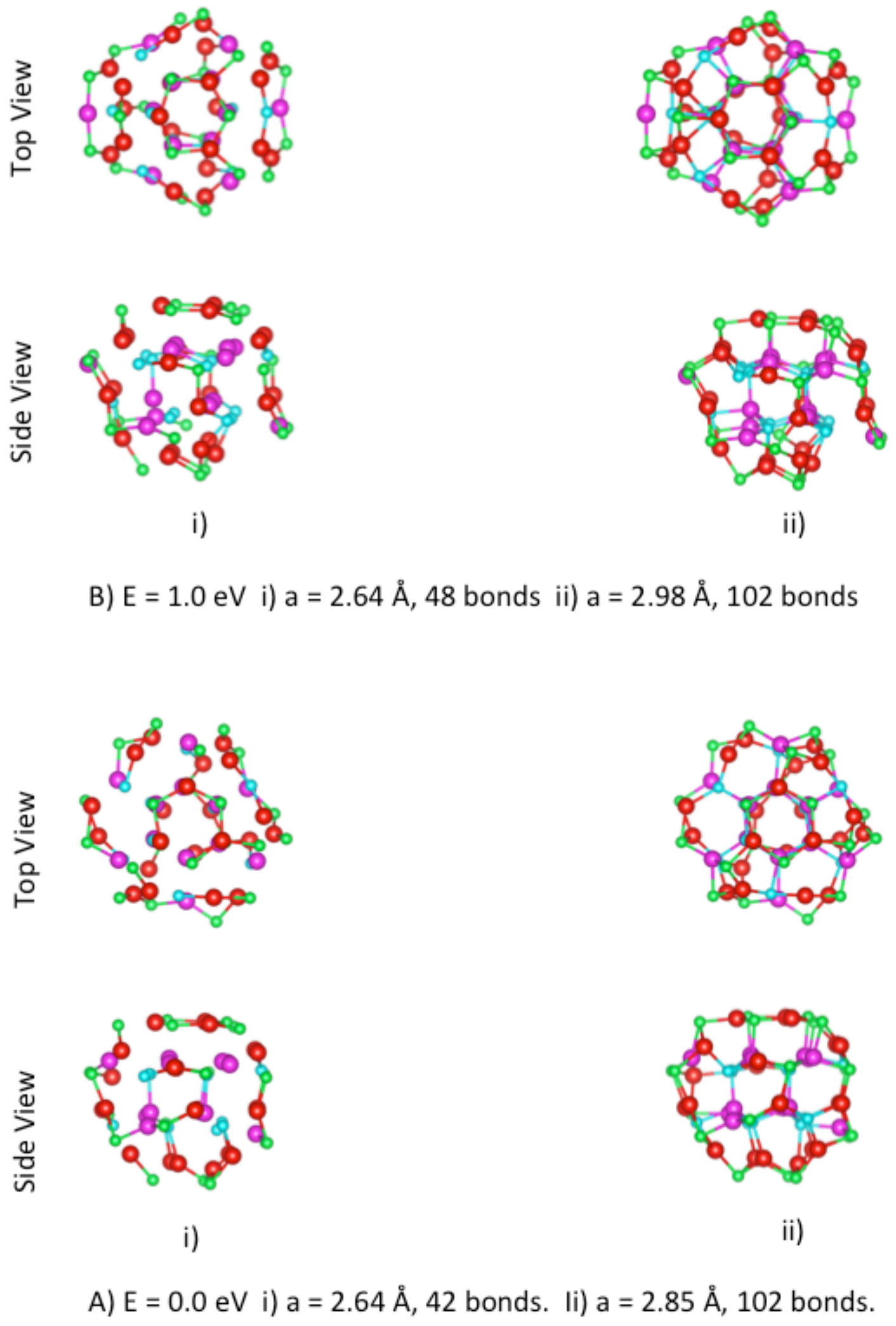

Fig. 3. $Cd_{33}Se_{33}$ Structures A (bottom rows) and B (upper rows)

In the top rows of Fig. 3, structure B, we present the results of optimization with constraint of symmetry. The effects of relaxation are in general the same as that of structure A except in one significant way. The constraint of symmetry has the effect of increasing (for Se) the bond distance along covalent bonding direction in order to reduce the strain. This effect is clearly seen at the lower right corner where the bonds are not formed at the prescribed bond length. Retention of the input symmetry of the structure, i.e. for instance smoothing out sharp edges, requires higher energy. The total energy of B is 1.0 eV above that of structure A.

Next we looked at surface charge distributions of the optimized structures as they are crucial for colloidal dispersion, transport through the solvent and eventually during the deposition to form thin films. In Fig. 4 we present the charge distributions of the structures next to each other. The bottom row corresponds to structure A and the top row corresponds to structure B. In both structures, the charge distributions clearly show how the Se atoms are pushed outward and the Cd atoms are pushed in. So the surface region is slightly more negative. Whether such local polarity is strong enough to cause repulsion among NCs suspended in solvents would be interesting to examine. The solvent has to be non-polar to prevent screening of the electric fields. We are unable to examine this at present since this is beyond the capability of the codes we used. The surface charge distribution also shows the ionic bonding, characteristic of II-VI semiconductors more prevalently. The reddish hues on doubly coordinated Se atoms are due to uncompensated electrons that were transferred from Cd. Some of the Cd atoms are deeper blue on the surface, indicative of more charge transfer. This effect is stronger in distorted sites at the edges, therefore suggesting how the geometric distortion can lead to more charge transfer than the usual transfer of the valence electrons. This again underscores the importance of the surface geometry and how it could lead to local charge imbalances. Structures with smoother charge distributions are preferable and again careful choice of the size of the NC is crucial for packing as well as transport through the colloid, and ultimately during deposition.

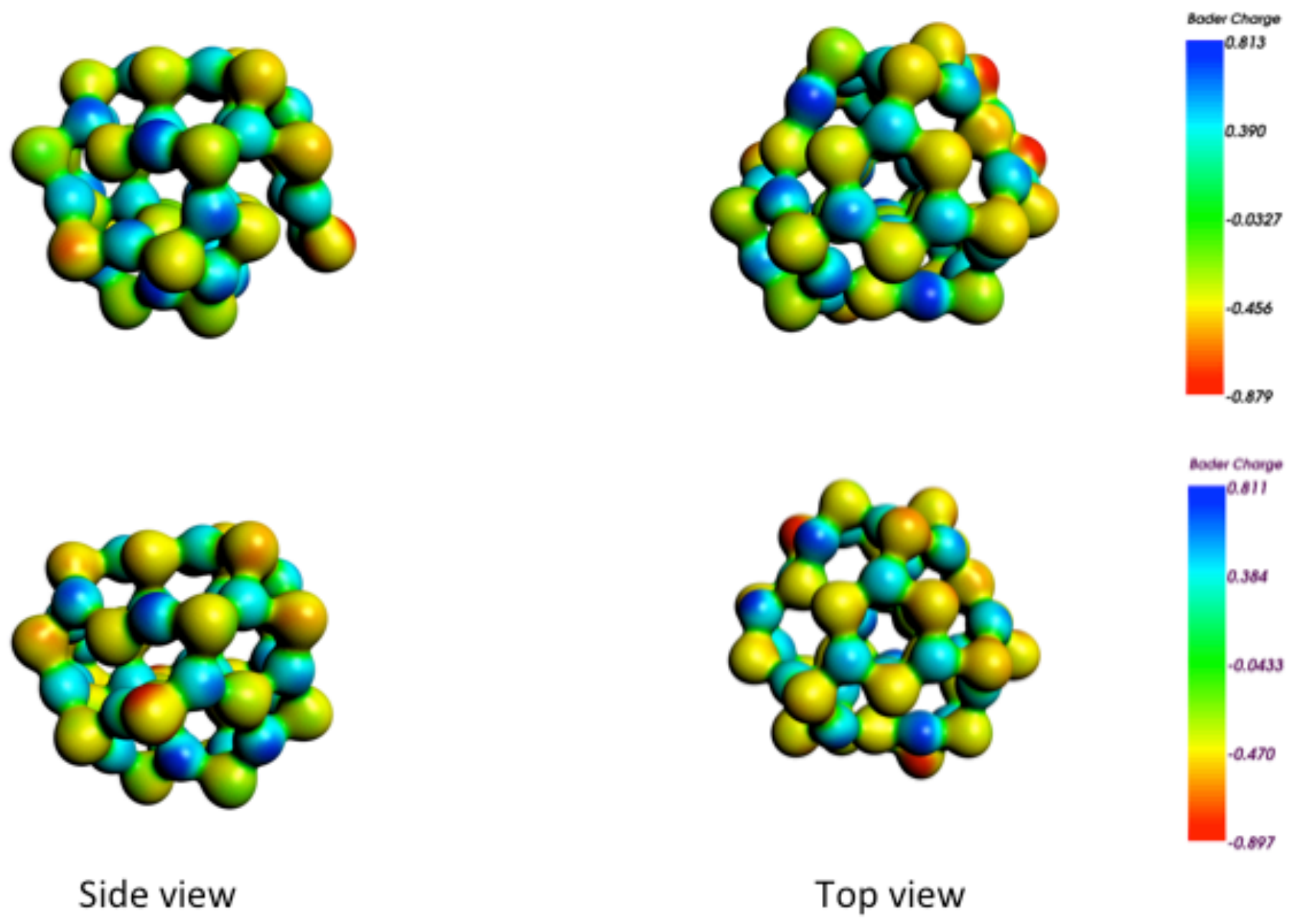

Fig. 4. Charge distributions for the optimized structures A (bottom row) and B (top row). The blue colors correspond to Cd and the yellowish green to Se sites.

The uneven distribution of the charges results in dipole and quadrupole moments. In Table 1, we give the dipole moments. The more smoothly relaxed structure of A that retains significantly the symmetry of the input structure shows the polarity along the z axis and approximate charge neutrality in the other two directions, as in the input structure. The retention of symmetry seems relevant to existence of well defined dipole moment, although one can not conclude with certainty from results of just $Cd_{33}Se_{33}$. Structure B involves symmetry operations that transform coordinates; the moment directions are not the same, therefore, the results are not presented for this structure as well as the other structures studied.

| | $P_x$ | $P_y$ | $P_z$ | $\|P\|$ (Debye) |
|---|---|---|---|---|
| A | 0.089 | 0.086 | -2.328 | 2.331 |

Table 1. Magnitude and components of Dipole moments of structure A

One can infer from the distributions of Cd and Se atoms of the input structure the existence of quadrupole moments. For instance on $(11\bar{2}0)$ surface, the net charge is approximately zero, but reflection about a plane going half way through the structure takes Cd atoms to Se atoms site and vice versa. From such geometry, it is seen that charges of opposite polarity exist at the end of the surface. The same kind of distribution of opposite polartity, but different magnitude exist on (1120) surface. Two of the off diagonal moments of A are of the same order as the diagonal elements. This is in contrast to results to a more symmetric distribution of charges of larger-sized NC as will be shown below.

| | $Q_{xx}$ | $Q_{yy}$ | $Q_{zz}$ | $Q_{xy}$ | $Q_{xz}$ | $Q_{yz}$ |
|---|---|---|---|---|---|---|
| A | 17.206 | -1.535 | -15.670 | -3.590 | -15.214 | 23.086 |

Table 2. Components of quadrupole moment of structure A in atomic units

Finally we looked at the charge distribution, rather the occupation of the electrons, the highest two occupied molecular orbitals. The electrons in these orbitals can be excited from HOMO to LUMO with electrostatic energy of the order of the HOMO-LUMO gap, creating local charge imbalance as seen in Fig. 5. Both HOMO and HOMO-1 orbitals are shown for each structure. The HOMO–LUMO gaps are 1.28 eV and 0.83 eV for the structures A and B respectively. The distorted structure B has narrower gaps than A. It is well known that DFT calculations underestimate the HUMO – LUMO gap, however, these numbers are consistent with the findings of Pudzer et al,[27] who attribute them to the formation of mid gap states formed between the HOMO and LUMO. We also find in agreement with previous works that the HOMOs reside mostly on Se sites on the edges as seen in Fig. 5. It is common experimentally to drive, for instance in electrophoretic deposition, the NCs to electrodes with an electric field. It is possible the electrostatic energy of the order of HOMO-LUMO gap or more would excite the electrons from the HOMO to LUMO, creating local charge imbalance. The negatively charged Se atoms would lose their excess electrons. This would reduce the overall negative polarity resulting from the puckering of Se atoms, as stated earlier.

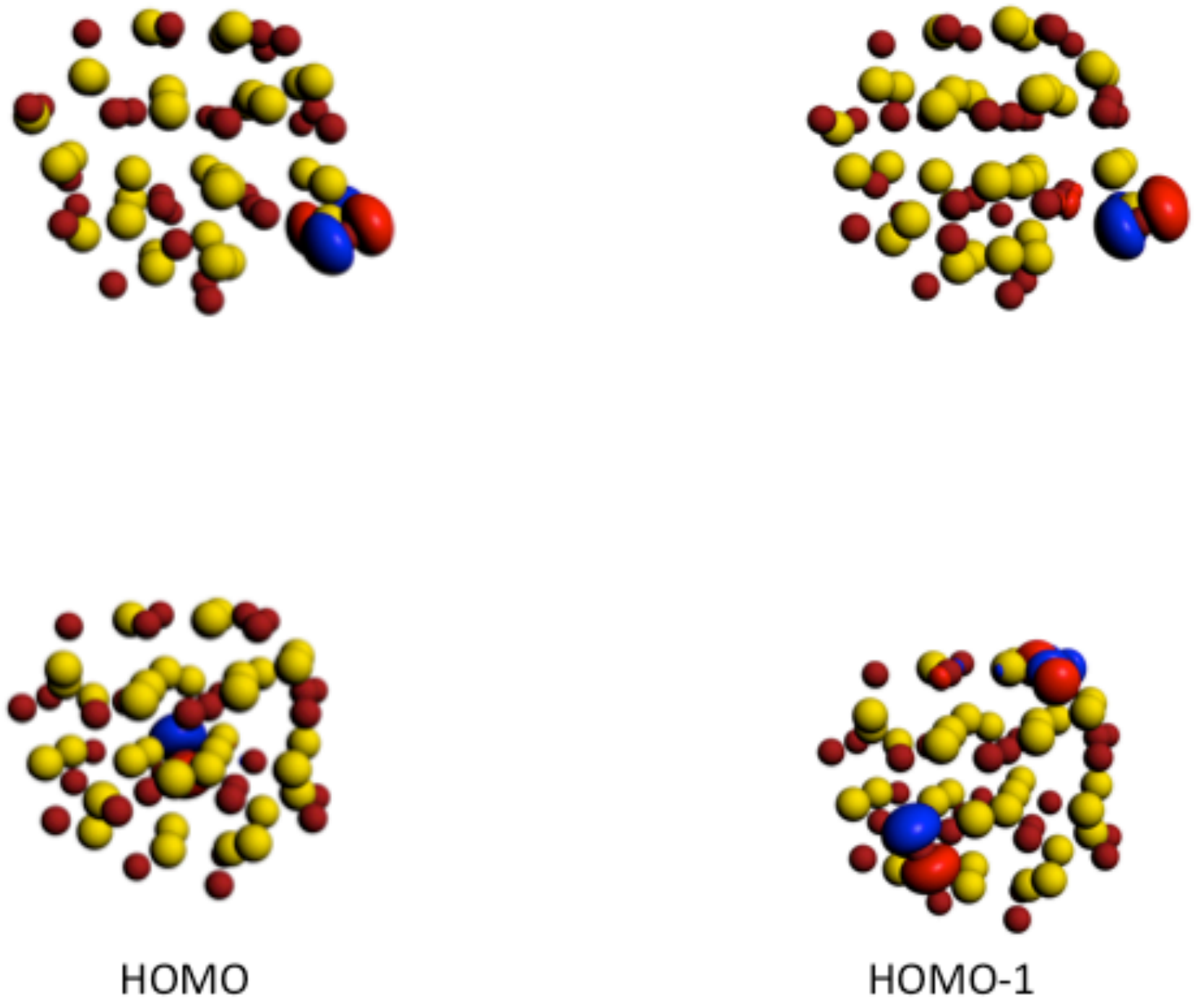

Fig. 5. Side views of the HOMO and HOMO-1 orbitals for A (bottom row) and B (top row)

**III-2. $Cd_{45}Se_{45}$**

In Fig. 6 the input geometry is shown; the coloring scheme and orientation are the same as in Fig. 1. First, the Cd-Se bond is set at the input bond length of 2.64 Å forming 144 bonds. The interior atoms make up 40 % of the total, 4 % points higher than that of $Cd_{33}Se_{33}$. Whether such a slight increase, or 24 atoms overall, will reduce the significant distortions seen in $Cd_{33}Se_{33}$, resulting from relaxation due finiteness in size will be examined here. The structure shows the same general features as $Cd_{33}Se_{33}$. The structure overall is more symmetric than $Cd_{33}Se_{33}$, for instance, as can be seen there is a reflection symmetry about a diagonal bisecting the side view in Fig. 6. The $(11\bar{2}0)$ and (1120) surfaces have identical layout of the surface atoms. Cd and Se sites are interchanged on the surfaces, however, run in reverse order and shifted in z direction with respect to each other. In general they are more symmetric than those of $Cd_{33}Se_{33}$.[29]

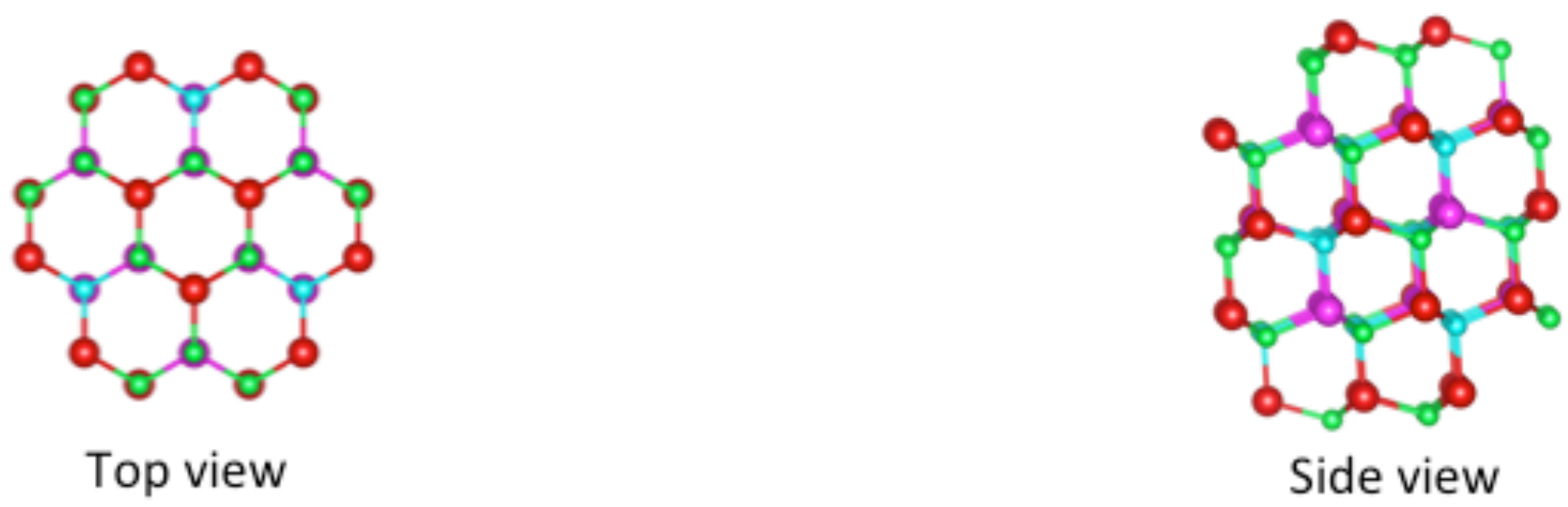

Fig. 6. $Cd_{45}Se_{45}$ input structure. Color scheme and orientation is as in Fig. 1.

The two optimizations performed on $Cd_{33}Se_{33}$ were repeated and the bond formations were analyzed using the Bader charge distributions to see if Cd – Cd bonds are formed. There were no Cd-Cd bonds formed contrary to the ball and stick models. In Fig. 7, we present the optimized structures for two different Cd – Se bond lengths for A and B respectively. The relaxation impacts the whole structure and not just the surface atoms, similar to that of $Cd_{33}Se_{33}$. Only 34.7% and 47.2%, of the input bonds have remained for A and B respectively and they form chains consisting almost entirely of surface atoms. Very few Cd-Se and Se-Cd-Se bonds are present. The relaxation has a significant effect on the atoms in the interior region as seen from non-bond formations in the interior at the input bond length a = 2.64 Å. Structure A recovers 95.1% of the input bonds at 2.82 Å, and 100 % at 2.95 Å. The distortions occur mainly at the corners, as seen in the side view of Fig. 7, and affect the interior region adjacent to them significantly. The view from the top shows 1) the existence of the distortions, and 2) how the Se atoms are pushed to the periphery.

Structure B, the top row, retains the symmetry of the input structure for the most part except at the lower right corner, where as in $Cd_{33}Se_{33}$ the Cd - Se distances have increased such that no bonds are formed. The (0001) surface is flat showing how Cd atoms have been pushed inwards so as to level off with the Se atoms that have been pushed outward. Its energy is 4.2 eV above that of A and its bond length increases to 3.4 Å to get the same number of bonds as in the input structure. This indicates that to get the bulk like crystal structure it is going to cost significant energy.

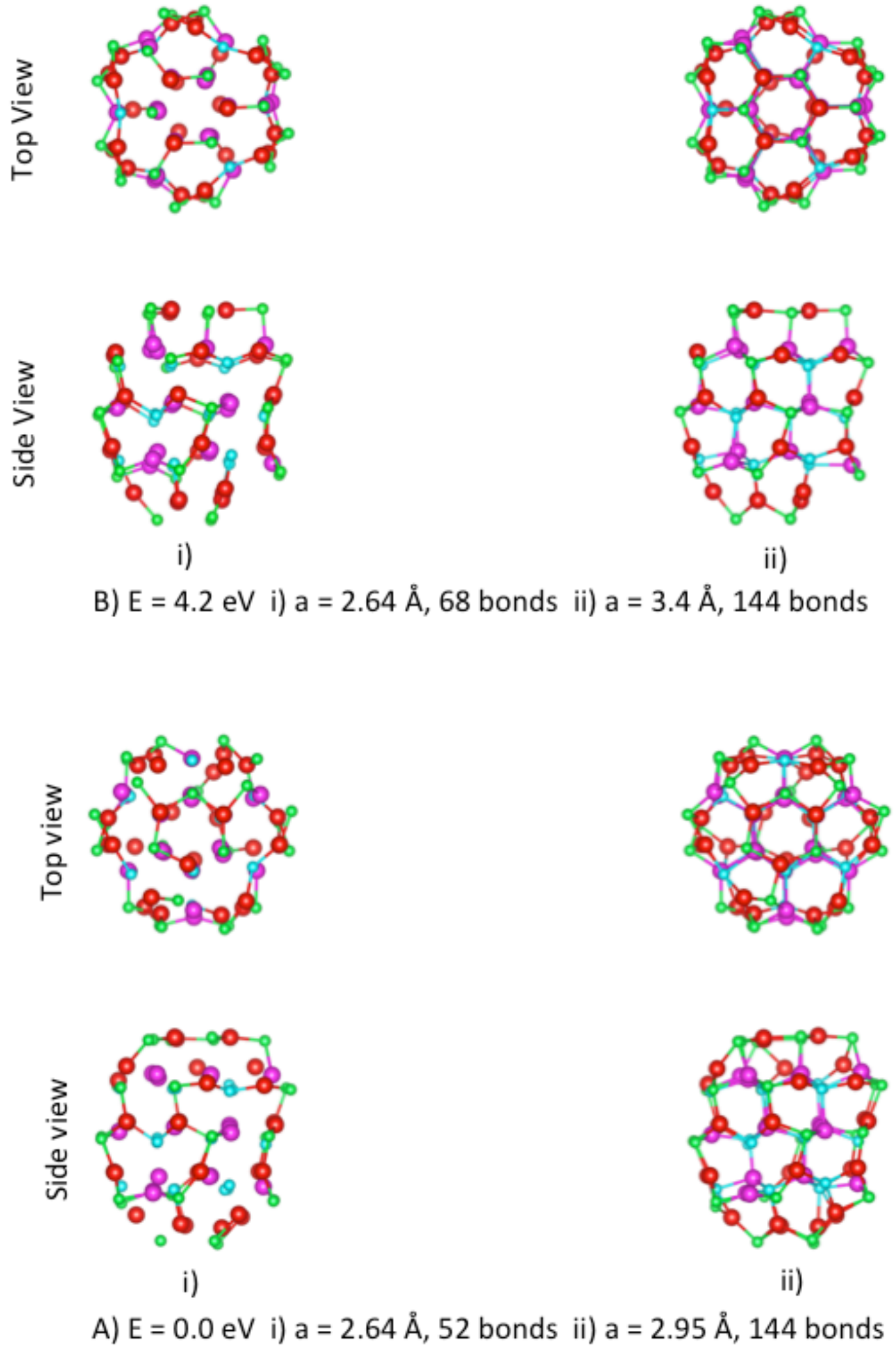

Fig. 7. $Cd_{45}Se_{45}$ structures A (bottom rows) and B (top rows)

The charge distributions are given in Fig. 8. Structure B is very distorted at the lower right hand corner of the side view. We note that as the system tries to preserve the tetrahedral geometry with the imposition of the symmetry constraint, it creates significant local charge imbalances as seen from the reddish and darker bluish sites. On the other hand, structure A with complete freedom to relax has much smoother charge distribution, except where the atoms are doubly coordinated.

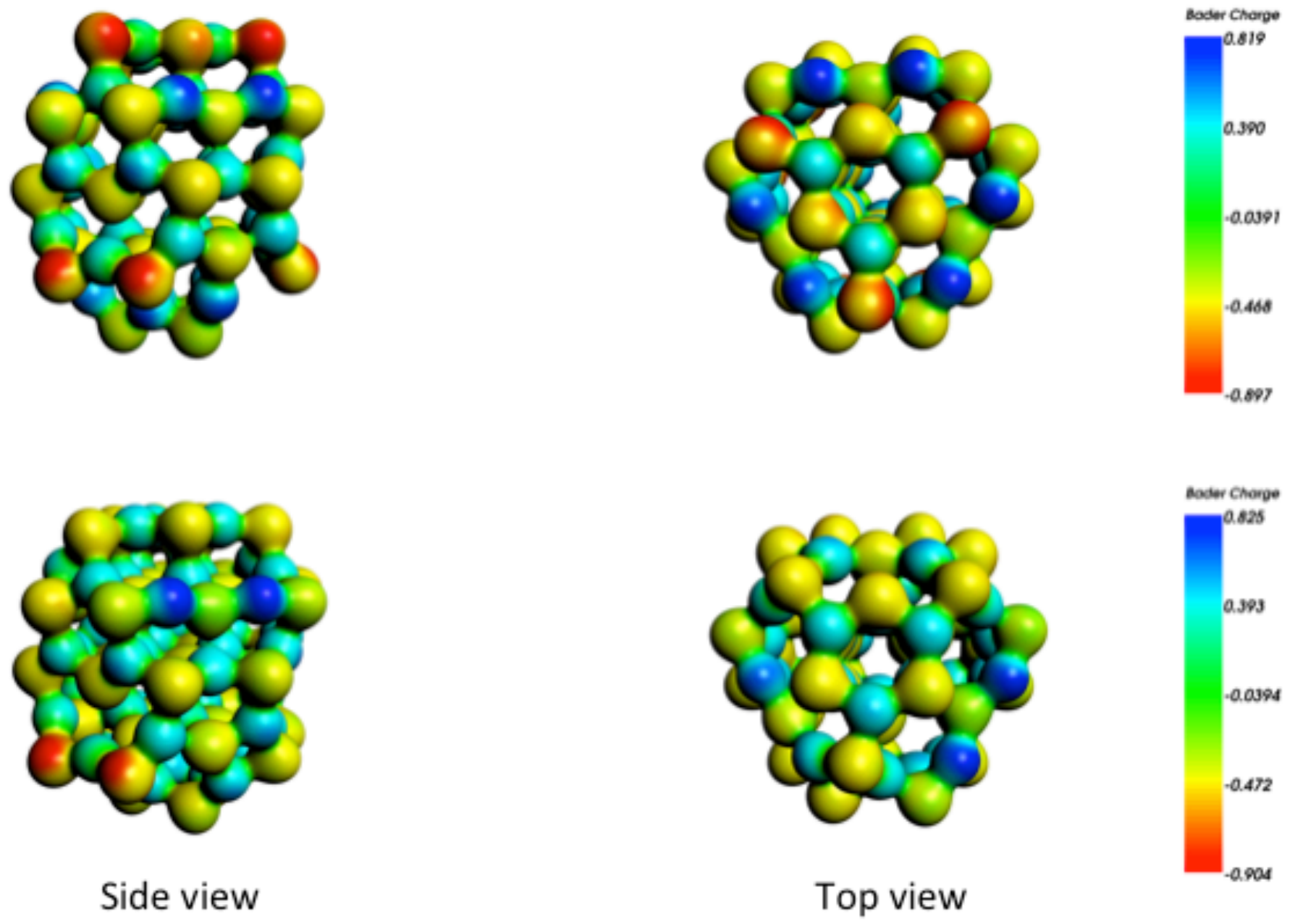

Fig. 8. Charge distributions for the optimized structures A (bottom row) and B (top row). The blue colors correspond to Cd and the yellowish green to Se sites.

The components of the dipole and quadrupole moments of structure A are given in Table 3. A sizable dipole component $P_x$ in addition to $P_z$ exists, indicative of a more symmetric distribution of the atoms on $(11\bar{2}0)$ and (1120) surfaces as discussed above. The quadrupole moments are traceless and there is one sizable off diagonal component corresponds partly to almost vanishing $P_y$ component. For a more symmetric distribution, as will be shown below for larger systems, the off diagonal terms are smaller by an order of magnitude.

| | $P_x$ | $P_y$ | $P_z$ | $\|P\|$ (Debye) |
|---|---|---|---|---|
| A | 3.955 | -0.444 | 2.286 | 4.590 |

| | $Q_{xx}$ | $Q_{yy}$ | $Q_{zz}$ | $Q_{xy}$ | $Q_{xz}$ | $Q_{yz}$ |
|---|---|---|---|---|---|---|
| A | 79.150 | -82.264 | 3.114 | -9.127 | -4.719 | -70.498 |

Table 3. The dipole and quadrupole moments of structure A

Finally we present the HOMO and HOMO-1 orbitals of structure A in Fig. 9. As seen in $Cd_{33}Se_{33}$, the orbitals are located at or near the periphery where the geometry is jagged. Both the HOMO and HOMO –1 are located at the same site. Any sizable electric field can strip off electrons form the locality and create local charge imbalance.

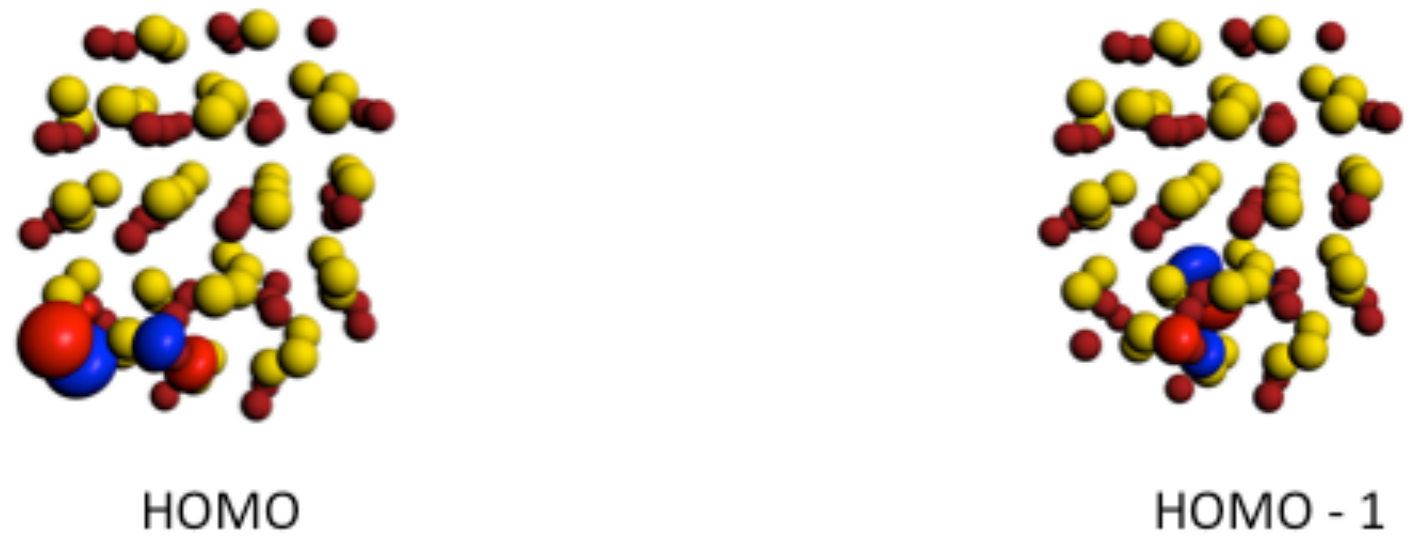

Fig. 9. Side views of the HOMO and HOMO-1 orbitals for structure A

### III-3. $Cd_{72}Se_{72}$

In Fig. 10, we present the $Cd_{72}Se_{72}$ input structure; it has 234 Cd-Se bonds. The top view shows that the periphery is surrounded by doubly bonded Cd and Se atoms. The polar surfaces consist of more atoms of Cd and Se than those of $Cd_{33}Se_{33}$ and $Cd_{45}Se_{45}$ and the number of Cd and Se atoms are equal unlike those of $Cd_{33}Se_{33}$ and $Cd_{45}Se_{45}$. The $(11\bar{2}0)$ and (1120) surfaces have the same kind of distribution as in $Cd_{45}Se_{45}$. The side view clearly shows that a reflection symmetry exists about a diagonal traversing from left to right, as in $Cd_{45}Se_{45}$. The percentage of interior atoms is 50 %, a 10 % increase from that of $Cd_{45}Se_{45}$.

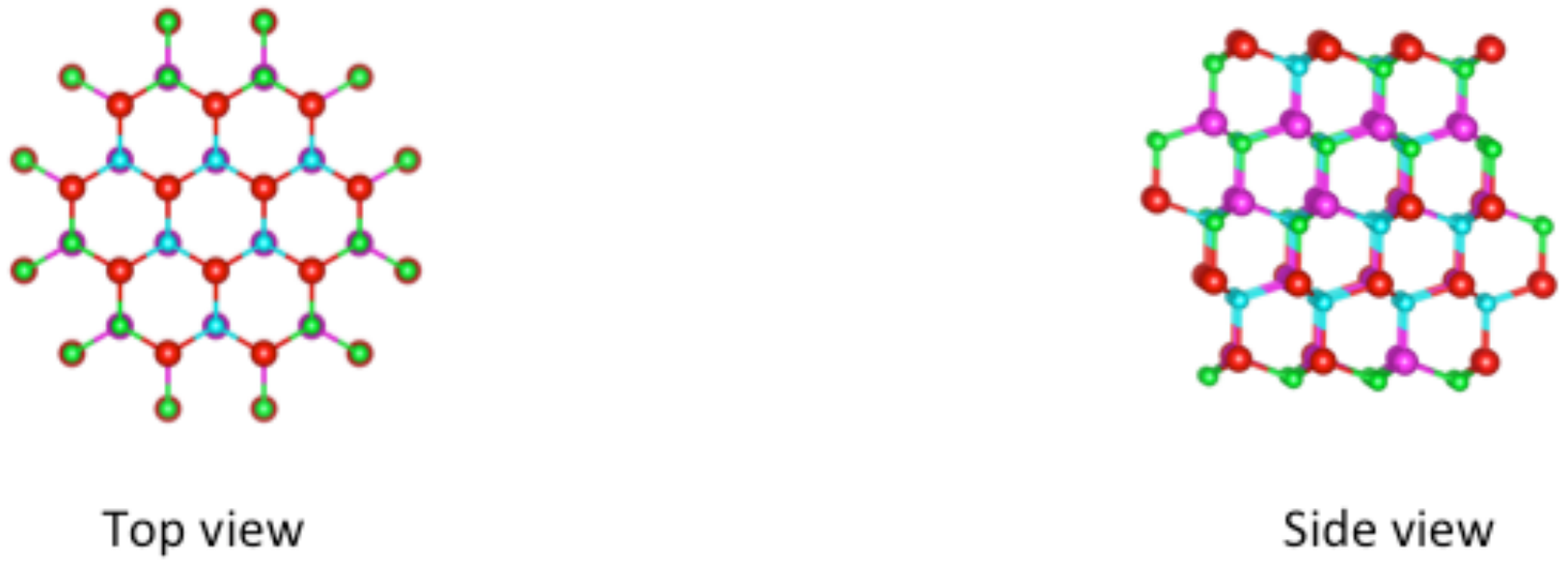

Fig. 10. $Cd_{72}Se_{72}$ input structure. Pink (Cd) and Blue (Se) are interior atoms.

The optimized structures A and B are given in Fig. 11. At the bond length of the input structure, we see that 39.7% and 38% of the bonds are left for A and B respectively. Unlike $Cd_{33}Se_{33}$ and $Cd_{45}Se_{45}$, the remaining bonds in structures A and B form shorter fragments instead of chains of atoms, except on the upper and lowermost polar surfaces. This indicates the robustness of the structure, that is, the relaxation has affected the positions of the atoms only in very minor way such that the effects or distortions are not large enough so as to make chains. With input number of bonds (or almost), structure A has minor distortions, and structure B has almost the same kind of geometry as the input, except around the edges. This shows that significant deviation has only occurred at the surface, which is the desired goal: to retain the interior of the atoms as intact as possible with their w*ü*rtzite structure. The energy difference between the two structures is much smaller than before, with A being .3 eV above B, or .024 eV per atom. This shows the size of the NCs has become big enough to withstand the forces that have distorted the tetrahedral geometry in the smaller structures of $Cd_{33}Se_{33}$ and $Cd_{45}Se_{45}$.

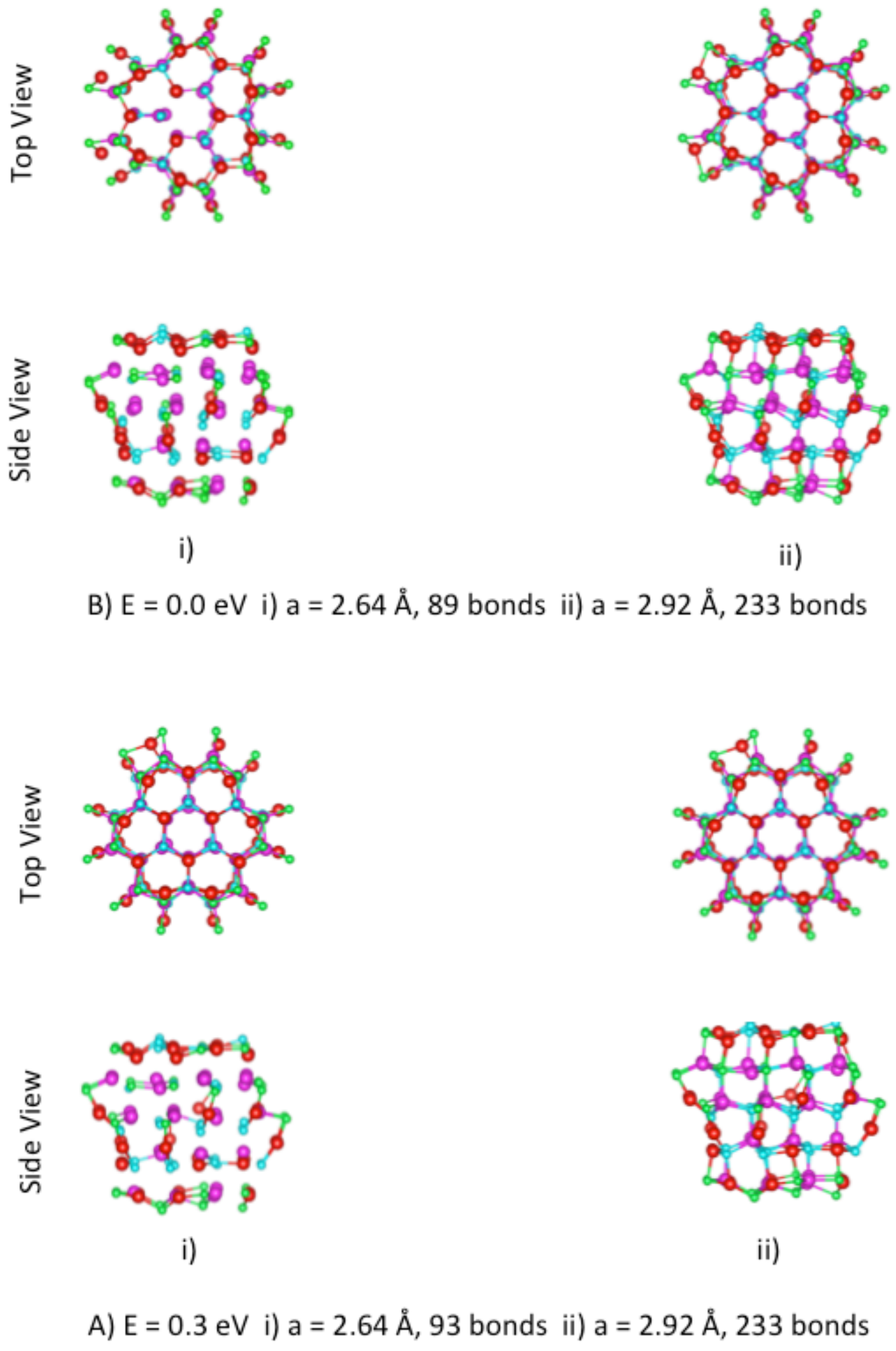

Fig. 11. $Cd_{72}Se_{72}$ structures A (bottom rows) and B (top rows)

The charge distributions associated with these structures are given in Fig. 12. The distributions of the two structures are almost identical. The doubly coordinated Se atoms exhbit reddish hues indicating more charge concentration. The Cd atoms on the surface are darker blue, indicating more charge tranfer to neighboring Se sites. Interestingly, these Cd atoms have three neighboring Se atoms. Overall, the surface is more ionic than those of $Cd_{33}Se_{33}$ and $Cd_{45}Se_{45}$.

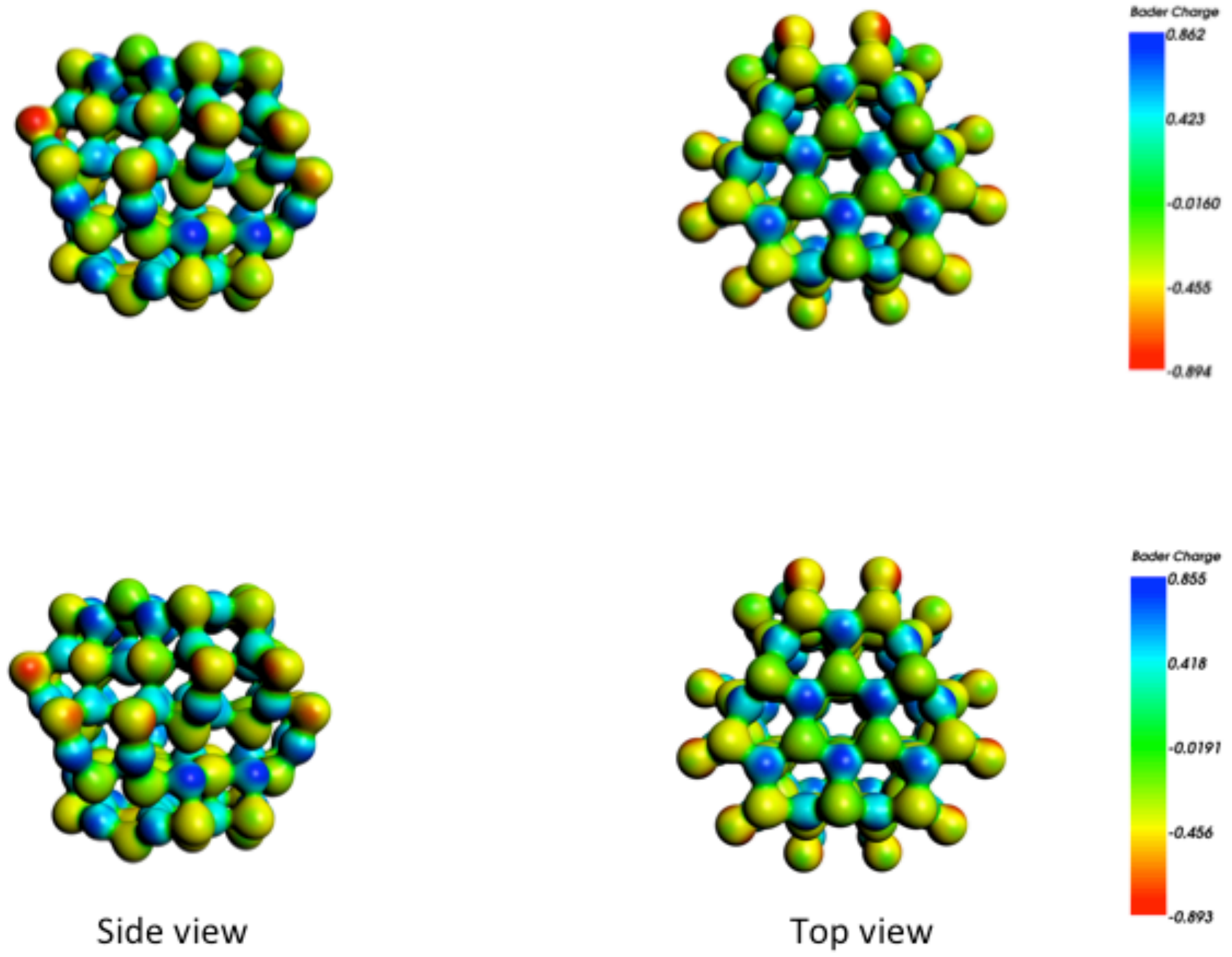

Fig. 12. Charge distributions for the optimized structures A (bottom row) and B (top row). The blue colors correspond to Cd and the yellowish green to Se sites.

The associated dipole and quadrupole moments of the charge distribution are given in Table 4. The dipole moment has decreased compared to previous structures. This is due to a more symmetric distribution of the charges, for instance, the number of Cd and Se atoms are equal on the polar (0001) and $(000\bar{1})$ for this structure unlike those of $Cd_{33}Se_{33}$ and $Cd_{45}Se_{45}$. The quadrupole moments for A are by orders of magnitude larger than the

| | $P_x$ | $P_y$ | $P_z$ | $\lvert P \rvert$ (Debye) |
|---|---|---|---|---|
| A | -0.770 | -0.649 | -1.718 | 1.991 |

| | $Q_{xx}$ | $Q_{yy}$ | $Q_{zz}$ | $Q_{xy}$ | $Q_{xz}$ | $Q_{yz}$ |
|---|---|---|---|---|---|---|
| A | -17.228 | -57.366 | 74.594 | -7.229 | 2.223 | -3.549 |

Table 4. The dipole and quadrupole moments of structure A off-diagonal ones. For ligated CdSe NCs, where the CdSe atoms are "locked" in position, Rabani also finds the same relationship in magnitudes between the diagonal and off diagonal components. This could be indicative of our geometry nearing that of the wurtzite geometry.

The HOMO and HOMO-1 orbitals are given in Fig. 13 below. The results are consistent with earlier results, that is, the orbitals are located on jagged sites on the surface.

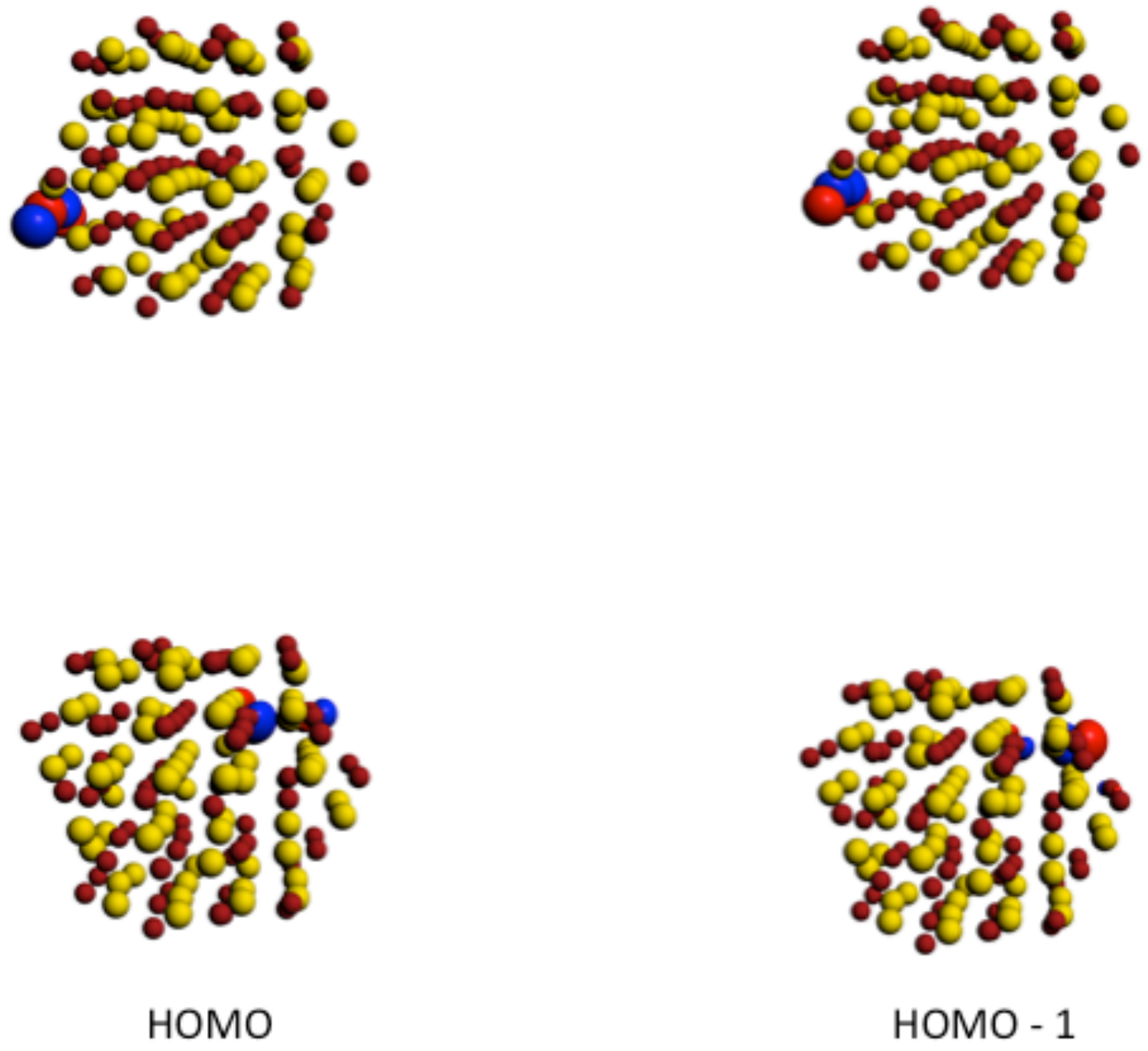

Fig. 13. Side views of the HOMO and HOMO-1 orbitals for structure A (bottom row) and structure B (top view)

### III-4. $Cd_{90}Se_{90}$

In Fig. 14 , we present the input geometry for $Cd_{90}Se_{90}$, which has 46.7% of the atoms as interior ones. Contrary to expectation, the percentage has gone down from that of $Cd_{72}Se_{72}$, which was 50 %. This shows that just a mere increase in the number of atoms is not sufficient to increase the atoms with tetrahedral coordination. In this case most of the atoms ended up triply bonded. So the right choice of the size of the NCs is very imortant to get a structure as close in geometry to würtzite structure as possible to reduce surface relaxation effects.

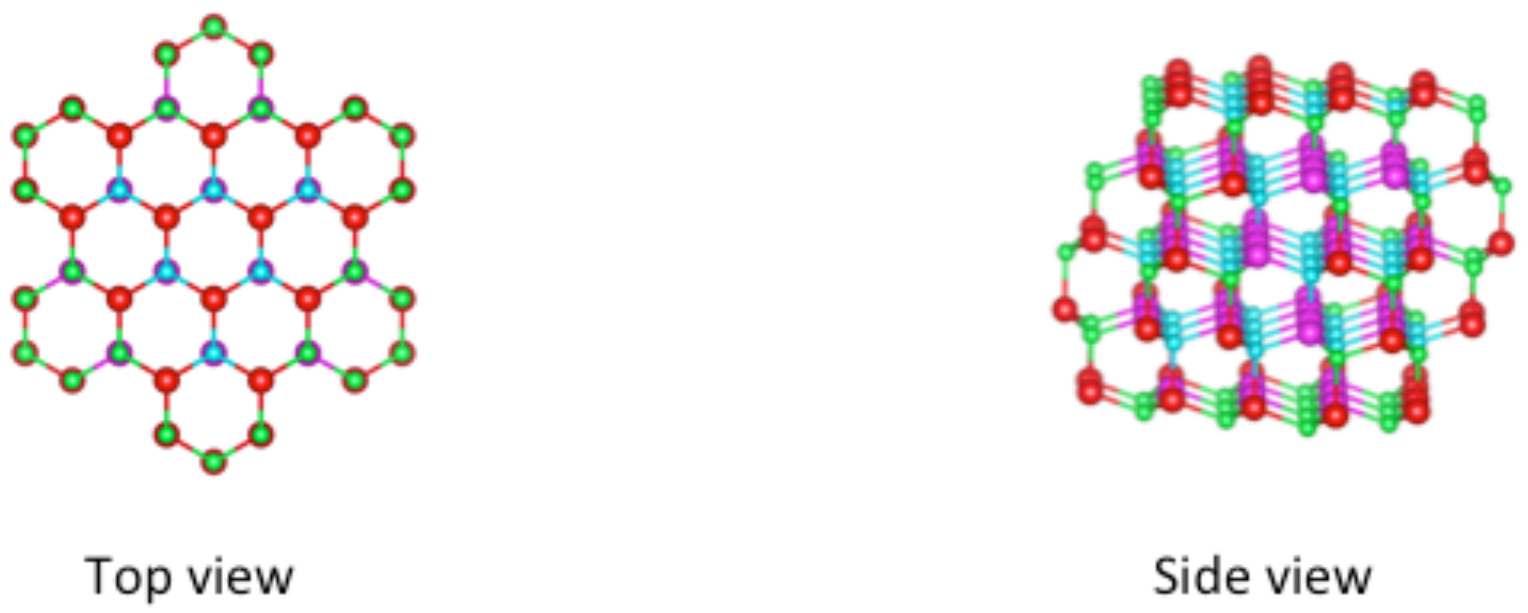

Fig. 14. $Cd_{90}Se_{90}$ input structure. Pink (Cd) and Blue (Se) are interior atoms

The results of the optimizationsof the structures A and B are given in Fig. 15. The first important outcome is that the energies of the two structures are the same. This clearly shows that the atoms are set in their symmetric geometry and are large enough in number that they withstand the forces causing significant distortions in the earlier smaller systems.

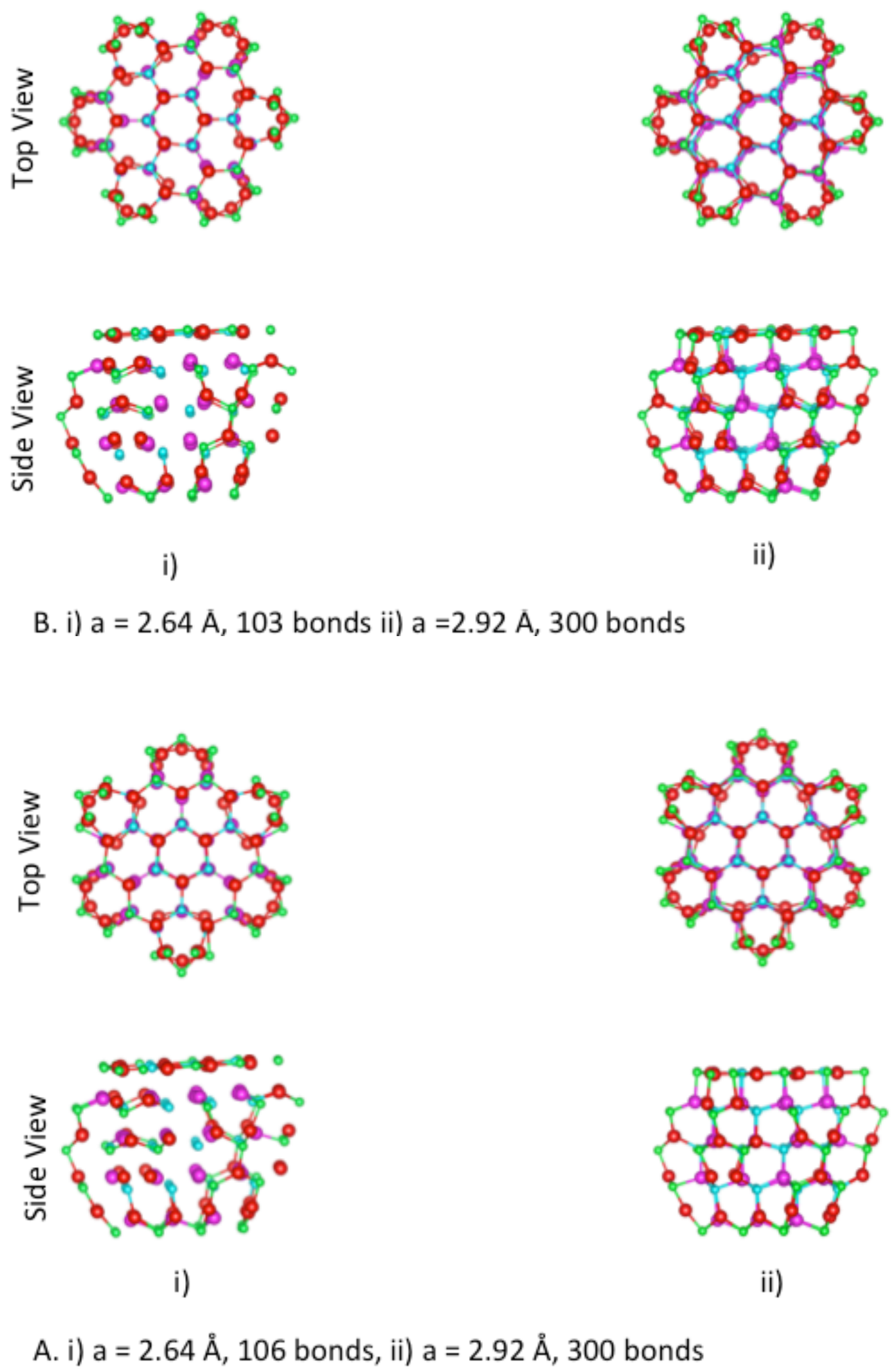

Fig. 15. $Cd_{90}Se_{90}$ structures A (bottom rows) and B (top rows). The total energies are equal.

The interior atoms retain their tetrahedral geometry and are well aligned. In fact, this has resulted in most of the surface atoms retaining their sites of the w*ü*rtzite geometry. The main effect of the relaxation is the flattening of the (0001) surface due to pulling in of Cd atoms and pulling out of Se atoms.

In Fig. 16 we present the charge distributions for both optimizations. They are almost identical; the side view shows in both cases that the charge distribution is quite smooth except at few Se surface sites with reddish hues. The covalent bonding in the interior region and the ionic nature of the surface is clearly seen.

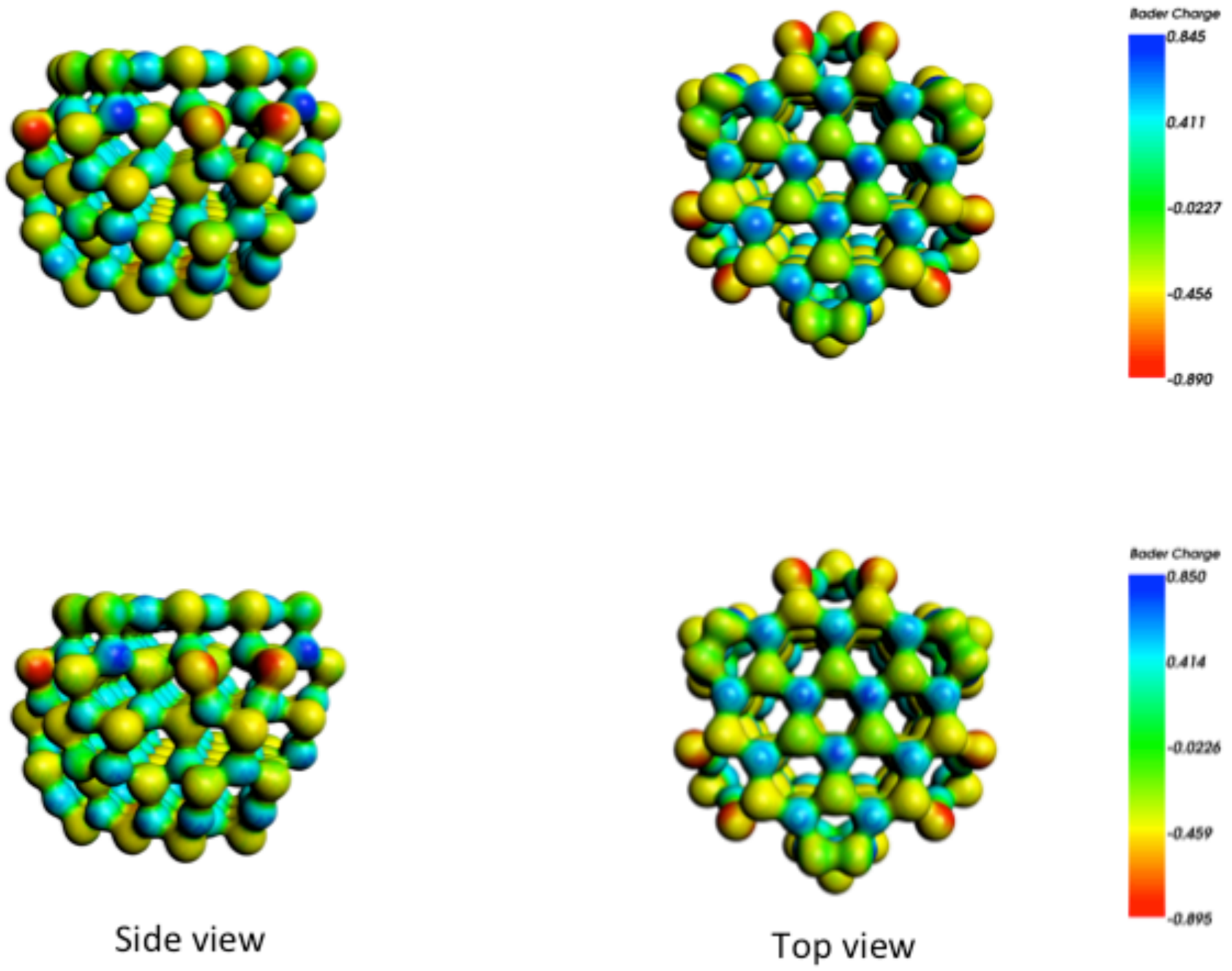

Fig. 16. Charge distributions for the optimized structures A (bottom row) and B (top row). The blue colors correspond to Cd and the yellowish green to Se sites.

The dipole and quadrupole moments of structure A are given in Table 5. The dipole moment has decreased significantly indicating a more symmetric charge distribution. The results follow the same pattern as in $Cd_{72}Se_{72}$, as the geometries are very similar. The quadrupole moments also show the same feature, i.e, the offdiagonal terms are by order of magnitude smaller as in $Cd_{72}Se_{72}$.

| | $P_x$ | $P_y$ | $P_z$ | $\|P\|$ (Debye) |
|---|---|---|---|---|
| A | 0.075 | 0.299 | 0.666 | 0.734 |

| | $Q_{xx}$ | $Q_{yy}$ | $Q_{zz}$ | $Q_{xy}$ | $Q_{xz}$ | $Q_{yz}$ |
|---|---|---|---|---|---|---|
| A | -35.963 | -31.632 | 67.595 | 2.194 | 3.779 | -2.022 |

Table 5. The dipole and quadrupole moments of structure A

In Fig. 17 we present the HOMO and HOMO-1 orbitals. Their location is different from all previous results where they are located on distorted or ragged edges. Here they have spread out across a plane one layer below the uppermost level. This could be a feature of a more symmetric structure, but we cannot conclude since it is just only one result.

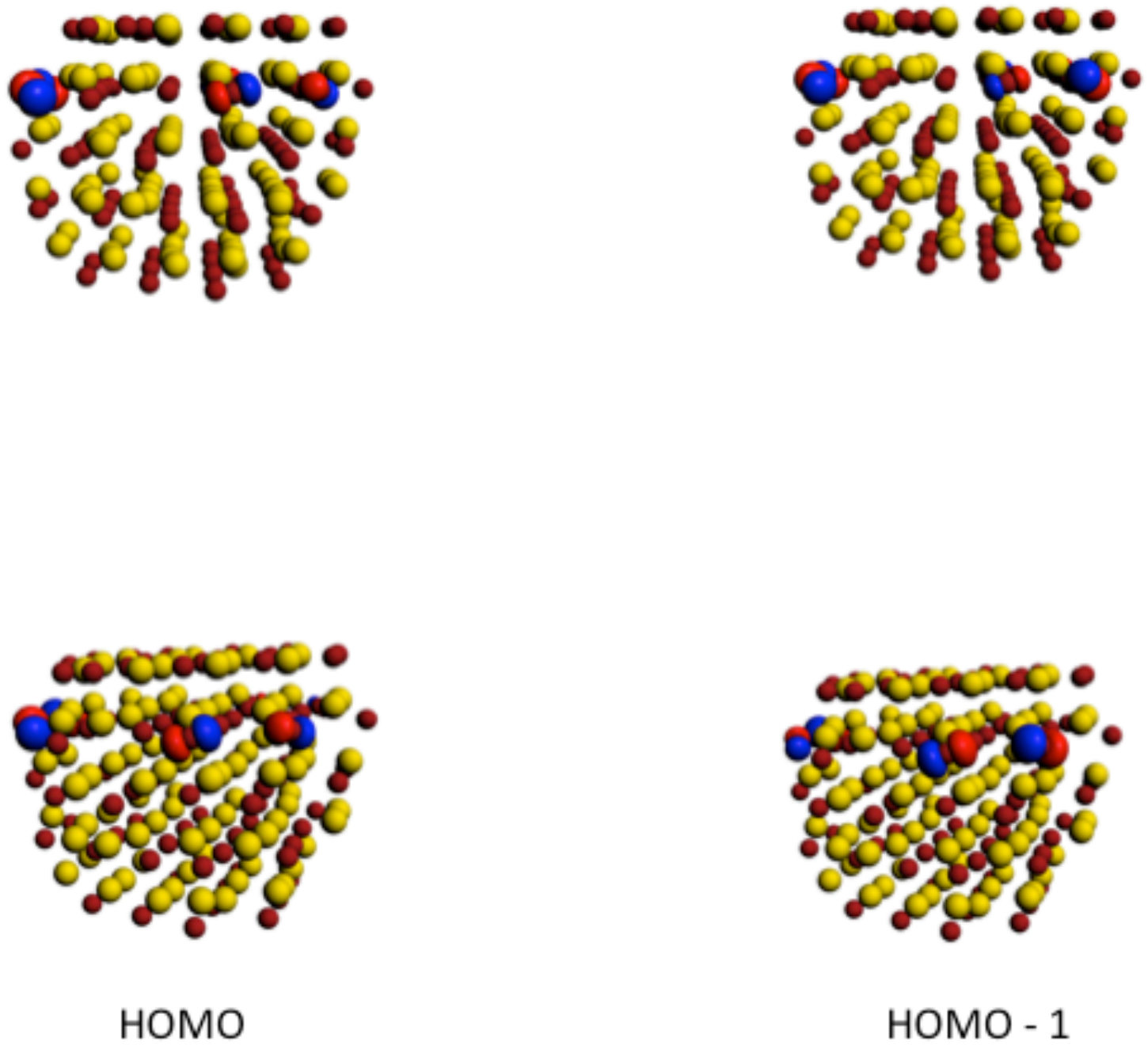

Fig. 17. Side views of the HOMO and HOMO-1 orbitals for structure A (bottom row) and structure B (top view)

## IV. Conclusion

In conclusion, we have investigated at four different structures of realistic sizes to determine the optimal structure and concomittant electrostatic properties for building macroscopic structure from microscopic units of nanosize. The NCs cut out from the w*ü*rtzite structure were subjected to two kinds of optimiztions, with or without symmetry constraint. The relaxation affects the whole structure, the interior as well as the surface atoms as seen from the percentage of input bonds remaining in the optimized structures. The samller systems of $Cd_{33}Se_{33}$ and $Cd_{45}Se_{45}$ are not robust enough to retain the input w*ü*rtzite geometry, i.e the structures are significantly distorted such that the unit cells are twisted and the surface have ragged edges. The concomittant charge distribuition has higher concentration at the distorted sites that consist mainly of doubly coordinated atoms. Such uneven distribution will affect the elecrtodynamical and electrodepositional properties involving NCs. The dipole and quadrupole moments of distorted structures are significant. On the other hand, the largest system studied of $Cd_{90}Se_{90}$ retains almost all of the input structure even though the relaxation affects all of the atoms. The relaxation significantly affects very few surface atoms that are doubly coordinated and distorts their bond structure. Otherwise the system is robust and withstands the optimizing forces and retains the w*ü*rtzite structure. $Cd_{72}Se_{72}$ is almost as robust as $Cd_{90}Se_{90}$. So one can infer with confidence that NCs of approximately 2 nm in diameter are the best candidates for macroscopic build up from microscopic units. They satisfy the two criteria of 1) small diameter and 2) retention of tetrahedral geometry as stated earlier.

**Acknowledgements**
This research is supported by the United States Army Research Office grant number W911NF-12-1-0047. *This research was also supported in part by the NSF through TeraGrid resources provided by TACC under grant number* TG-DMR100055.

[1] D. V. Talapin, J.-S. Lee, M. V. Kovalenko, and E. V. Shevchenko, Chem. Rev. **110**, 389 (2010)
[2] C. B. Murray, D. J. Norris, M. G. Bawendi, J. Am. Chem. Soc. **115**, 8706 (1993)
[3] P. Yang, S. Tretiak, S. Ivanov, J Clust Sci **22**, 405 (2011)
[4] L. R. Becerra, C. B. Murray, G. G. Griffin, and M. G. Bavendi J. Chem. Phys. **100**, 3297 (1994)
[5] J. E. B. Katari, V. L. Colven, and A. P. Alivistos, J. Phys. Chem. **98**, 4109 (1994)
[6] S. A. Fischer, A. M. Crotty, S.V. Kilina, S. A. Ivanov and S. Tretiak, Nanoscale **4**, 904 (2011)
[7] N. Y. Morgan, C. A. Leatherdale, M. Drndiae, M. V. Jarosz, M. A. Kastner, M. G. Bawendi, Phys. ReV. B **66**, 075339 (2002)
[8] D. V. Talapin, C. B. Murray, Science **310**, 86 (2005)
[9] M. S. Kang, A. Sahu, D. J. Norris, C. D. Frisbie, Nano Lett. **10**, 3727 (2010)
[10] K. P. Kadlag, M. J. Rao, and A. Nag, J. Phys. Chem. Lett. **4**, 1676 (2013)
[11] Z. C. Holman and U. R. Kortshagen, Nano Lett. **11**, 2133 (2011)
[12] X. Lin, C. Wang, S. Xu and Y. Cui, J. Phys. D: Appl. Phys. **47**, 015103 (2014)
[13] E. Rabani , B. Hetényi , B. J. Berne , and L. E. Brus, J. Chem. Phys. **110**, 5355 (1999)
[14] E. Rabani, J. Chem. Phys.**115**, 1493 (2001)
[15] P. Deglmann, R. Ahlrichs, and K. Tsereteli, J. Chem. Phys. **116**, 1585 (2002)
[16] G.L. Gutsev, M.D. Mochena, B.C. Saha, C.A.Weatherford, P.A. Derosa, J. Comput. Thoret . Nanosci. **7**, 254 (2010)
[17] We used the ADF software to cut out the spheres. For $Cd_{33}Se_{33}$, we cut out $Cd_{36}Se_{36}$ and removed 3 atoms each so to get $Cd_{33}Se_{33}$ that was used in a number of studies.
[18] G. Kresse and J. Hafner, J. Phys.: Condens. Matt. **6**, 8245 (1994)
[19] G. Kresse and J. Hafner, Phys. Rev. B **47**, 558 (1993)
[20] G. Kresse and J. Furthmüller, Phys. Rev. B **54**, 11169 (1996)
[21] J. P. Pedrew and Y. Wang, Phys. Rev. B 45, 13244 (1992)
[22] J. J. P. Stewart, J. Mol. Modeling **19**, 1 (2013)
[23] G. te Velde, F.M. Bickelhaupt, S.J.A. van Gisbergen, C. Fonseca Guerra, E.J. Baerends, J.G. Snijders and T. Ziegler, J. Comput. Chem. **22**, 931 (2001)
[24] K. Momma and F. Izumi, Cryst. Comput., IUCr Newslett. 7, 106 (2006)
[25] A. Kasyuya et al, Nat. Mater. 3, 99 (2004)
[26] We plotted the three dimensional balls instead of contour plots because the atoms are not aligned with each other, therefore, the contour plots as two dimensional projection can not give a correct picture of the structure.
[27] A. Puzder, A. J. Williamson, F. Gygi, G. Galli, Phys. Rev. Lett. 92, 217401 (2004)
[28] V. Albert, S. A. Ivanov, S. Tretiak, S. V. Kilina, J. Phys. Chem. C 115, 15793 (2011)
[29] The $Cd_{45}Se_{45}$ input was cut out by the ADF software as is. On the other hand, the $Cd_{33}Se_{33}$ input was derived from the cut-out $Cd_{36}Se_{36}$ structure by stripping off 3 atoms of Cd and Se respectively.